\documentclass{PoS_oldRef}
\usepackage{graphicx}
\usepackage{graphics}

\title{Cluster Radio Halos at the crossroads between astrophysics and cosmology in the SKA era}

\ShortTitle{Radio Halos and SKA}

\author{
\speaker{R. Cassano}$^1$,
G. Bernardi$^2$, 
G. Brunetti$^1$, 
M. Br\"uggen$^3$, 
T. Clarke$^4$, 
D. Dallacasa$^{1,5}$, 
K. Dolag$^6$, 
S. Ettori$^7$, 
S. Giacintucci$^8$, 
C. Giocoli$^5$, 
M. Gitti$^{1,5}$, 
M. Johnston-Hollitt$^9$, 
R. Kale$^{1,10}$, 
M. Markevitch$^{11}$, 
R. Norris$^{12}$, 
M. Pandey-Pommier$^{13}$, 
G.W. Pratt$^{14}$, 
H. R\"ottgering$^{15}$, 
T. Venturi$^1$
\\
$^1$ INAF-IRA, Bologna, IT; $^2$ SKA SA, Rhodes University, SA ; $^3$ Hamburg University, Germany; $^4$ NRL, US; $^5$ DIFA, University of Bologna, IT; $^6$ USM, DE; $^7$ INAF - OAB, IT; $^8$ University of Maryland, US; $^9$ University of Wellington, NZ; $^{10}$ NCRA, Pune, IN; $^{11}$ NASA/GSFC, US; $^{12}$ CSIRO, AU; $^{13}$ CRAL, Lyon, FR; $^{14}$ CEA Saclay - IRFU, Service d' Astrophysique, FR; $^{15}$ Leiden Observatory, NL
\\
 E-mail: \email{rcassano@ira.inaf.it}
 }

\abstract{Giant Radio Halos ({\bf RH}) are diffuse, Mpc-sized, synchrotron radio sources observed in a fraction of galaxy clusters. They probe the energy content and properties of relativistic particles and magnetic fields in galaxy clusters and their imprint on cluster formation and evolution. RHs are found in merging clusters, suggesting that they are generated as a result of the dissipation of gravitational energy during the hierarchical sequence of mergers that leads to the formation of clusters themselves.
The current leading scenario for the origin of RHs assumes that turbulence generated during cluster mergers re-accelerates pre-existing fossil and/or secondary electrons in the intra-cluster-medium (ICM) to the energies necessary to produce the observed radio emission. Moreover, more relaxed clusters could host diffuse ``off state'' halos, fainter than classical RHs, produced by secondary electrons. In this Chapter we use Monte Carlo simulations, that combine turbulent-acceleration physics and the generation of secondaries in the ICM, to calculate the occurrence of RHs in the Universe, their spectral properties and connection with properties of the hosting clusters at different cosmic epochs.
Predictions for SKA1 surveys are presented at low (100-300 MHz) and mid (1-2 GHz) frequencies assuming the expected sensitivities and spatial resolutions of SKA1. SKA1 will step into an unexplored  territory allowing us to study the formation and evolution of RHs in a totally new range of cluster masses and redshift. Based on our study, SKA1 observations will allow firm tests of the current theoretical hypothesis. In particular we show that the combination of SKA1-LOW and SUR will allow the discovery of $\sim 1000$ ultra-steep-spectrum halos and to detect for the very first time ``off state'' RHs. We expect that at least $\sim2500$ giant RHs will be discovered by SKA1-LOW surveys up to $z\sim 0.6$. Remarkably these surveys will be sensitive to RHs in a cluster mass range (down to $\sim 10^{14}$M$_{\odot}$) and redshifts (up to $\sim1$) that are unexplored by current observations. SKA1 surveys will be highly competitive with present and future SZ-surveys in the detection of high-redshift massive objects.}

\FullConference{
Advancing Astrophysics with the Square Kilometre Array\\
June 8-13, 2014\\
Giardini Naxos, Italy}

\newcommand{\skipthis}[1]{}

\newcommand\apj{ApJ}

\def\ltsim{\; \raise0.3ex\hbox{$<$\kern-0.75em \raise-1.1ex\hbox{$\sim$}}\; }
\def\gtsim{\; \raise0.3ex\hbox{$>$\kern-0.75em \raise-1.1ex\hbox{$\sim$}}\; }
\def\ie{{\it i.e.,~}}
\def\ltapprox{\raise 2pt \hbox {$<$} \kern-1.1em \lower 5pt \hbox {$\approx$}}

\def\eg{{\it e.g.,~}}

\begin{document}

\section{Introduction}
Giant Radio Halos (RHs) are diffuse, Mpc-sized, synchrotron radio sources with steep radio spectra ($\alpha>1$, with $f(\nu)\propto \nu^{-\alpha}$) that are observed in the central regions of a fraction of galaxy clusters (\eg Feretti et al. 2012, for a review of the observational properties). They probe the energy content and properties of relativistic particles and magnetic fields in galaxy clusters and their imprint on cluster formation and evolution. RHs are always found in merging clusters\footnote{An exception is the halo in the cool-core cluster CL1821+643 (Bonafede et al 2014). One hypothesis suggested by these authors is however that this cluster is undergoing a minor or off-axis merger.}, suggesting that they are generated as a result of the dissipation of gravitational energy during the hierarchical sequence of mergers that leads to the formation of clusters themselves (\eg Brunetti \& Jones 2014, for a recent review).
A popular idea for the origin of RHs is based on the hypothesis that turbulence generated during cluster mergers re-accelerates pre-existing fossil and/or secondary electrons in the intra-cluster-medium (ICM) to the energies necessary to produce the observed radio emission (\eg Brunetti et al. 2001; Petrosian 2001). 
According to turbulence-acceleration models the formation history of RHs depends on the cluster merging rate throughout cosmic epochs and on the
mass of the hosting clusters themselves, which ultimately sets the energy budget that is available for the acceleration of relativistic particles. A key expectation is that RHs should preferentially be found in massive objects undergoing energetic merging events, whereas  they should be rarer in less massive merging-systems. The reason making RHs less common in smaller systems is that the acceleration mechanism in the case of less energetic mergers is expected to generate increasingly steep spectra which become under luminous at higher frequencies (\eg Cassano et al. 2006, 2010a). This theoretical conjecture is consistent with the observed radio bimodality in galaxy clusters and its connection to cluster dynamics (\eg Cassano et al. 2010b), it is also supported by the discovery of RHs with very-steep spectrum (\eg Brunetti et al. 2008; Dallacasa et al. 2009). 

The generation of secondary particles due to inelastic collisions between relativistic protons and thermal protons in the ICM is another mechanism for the generation of cluster-scale diffuse emission (\eg Dennison 1980; Blasi \& Colafrancesco 1999). Studies in the radio and gamma-rays suggest that the contribution to RHs due to this latter mechanism is sub-dominant (Ackermann et al. 2010; Brunetti et al. 2012), however in more relaxed clusters it is expected to produce diffuse radio sources, called ``off state'' halos, fainter than classical RHs (\eg Brunetti \& Lazarian 2011; Brown et al. 2011; Donnert et al. 2013). 

In this Chapter we will adopt the theoretical framework described above and use Monte Carlo simulations, that combine turbulent-acceleration physics and the generation of secondaries in the ICM, to calculate the occurrence of RHs in the Universe and their spectral properties. These simulations provide a physically motivated way to model the connection between RHs (and their characteristics) and the thermodynamical properties and mass of the hosting clusters at different cosmic epochs. Predictions for SKA1 surveys are presented at low (100-300 MHz) and mid (1-2 GHz) frequencies assuming the expected sensitivities and spatial resolutions of SKA1 (SKA1 Performance Memo by R. Braun).

\noindent A $\Lambda$CDM cosmology ($H_{o}=70\,\rm km\,\rm s^{-1}\,\rm Mpc^{-1}$, $\Omega_{m}=0.3$, $\Omega_{\Lambda}=0.7$) is adopted.


\section{Statistical modeling of diffuse radio emission in clusters: "turbulent" and "off-state" halos}

\begin{figure}
\begin{center}
\includegraphics[width=6cm,height=6cm]{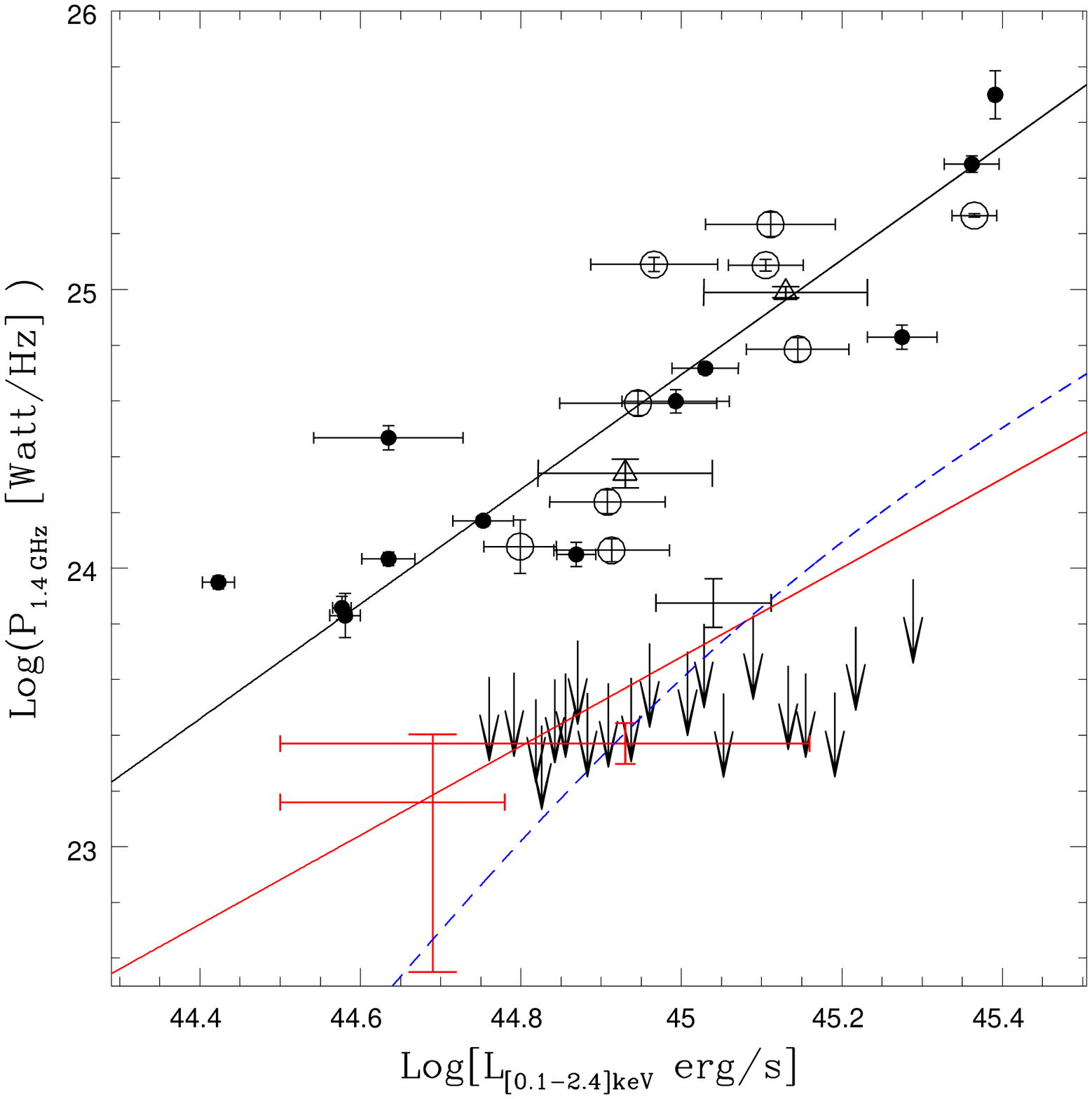}
\includegraphics[width=6cm,height=6cm]{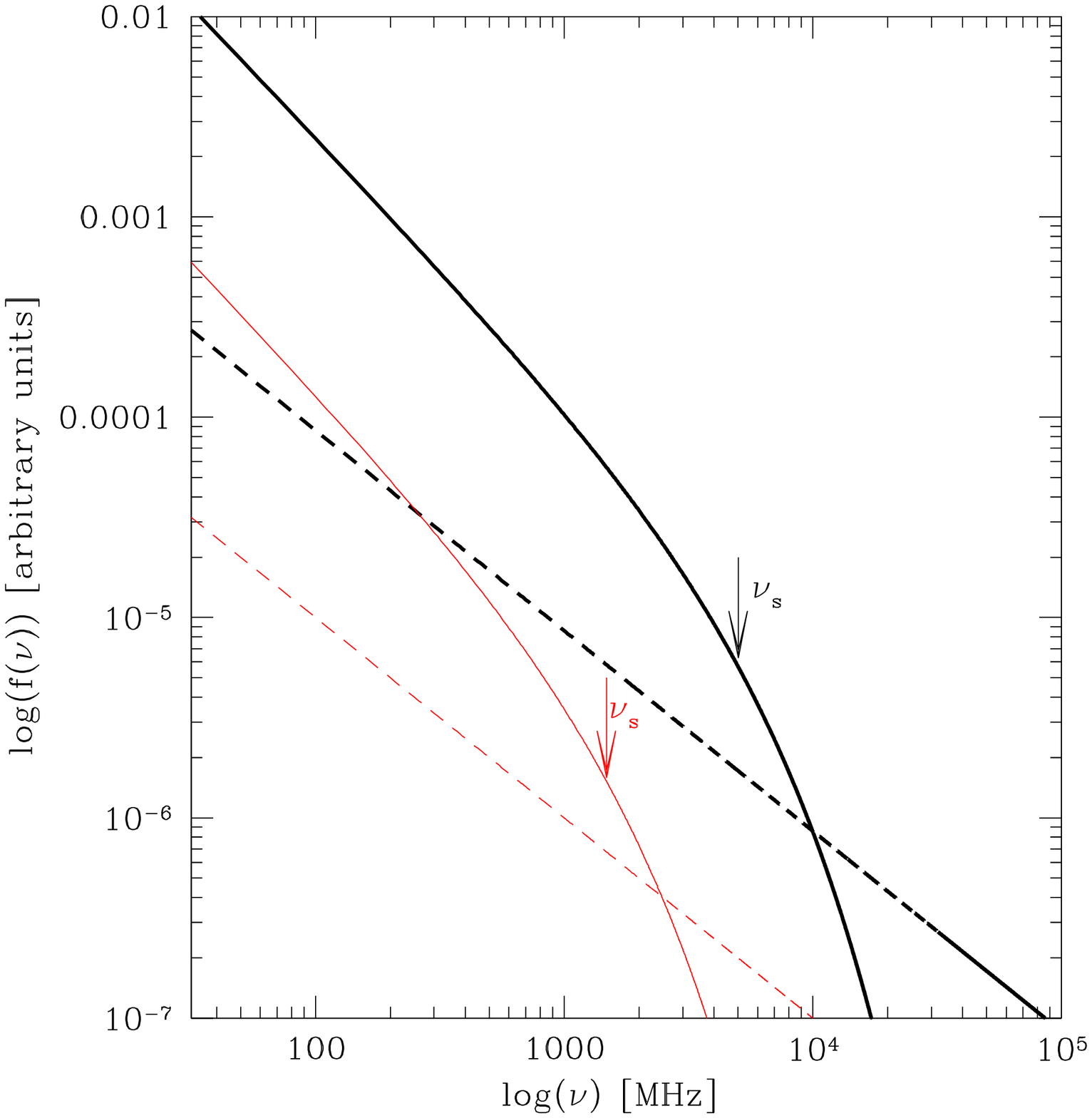}
\caption{{\it Left Panel} Distribution of clusters in the $P_{1.4}-L_X$ plane (from Brunetti et al. 2009). Clusters from the literature (black filled circles) and clusters belonging to the ``GMRT RH Survey'' (open circles and arrows) are reported. The red crosses are obtained by staking the radio images of clusters from the SUMSS survey (Brown et al. 2011). On the same figure we also report the scalings adopted here for halos produced by secondary electrons (red solid line;see text for details). 
{\it Right Panel}: Reference spectra of ``turbulent'' RHs (solid lines) and ``off-state'' (hadronic) halos (dashed lines) in a massive 
(\ie $M_v\sim 2.5\times 10^{15}\, M_{\odot}$; black thick lines) and less massive (\ie $M_v\sim 10^{15}\, M_{\odot}$; red thin lines) cluster. 
Arrows indicate the position of the steepening frequency, $\nu_s$, in the two cases. The turbulent spectra are computed assuming
in both cases a merger event with a sub-clump of mass $\Delta M=5\times 10^{14}\,M_{\odot}$ at $z=0.023$.}
\label{Fig.Lx_P_sec}
\end{center}
\end{figure}

A detailed description of the theoretical-statistical model that we will use in this Chapter can be found in Cassano \& Brunetti (2005) and Cassano et al. (2006), while applications to RH predictions for future surveys (with LOFAR, Apertif on WSRT, and ASKAP) can be found in Cassano et al. (2010a, 2012). 
In this Sect. we provide a summary of the theoretical framework and of the most important implications for RH statistical properties and connection with the host clusters.
In this contribution we follow the formation and evolution of two populations of cluster RHs: {\it (i)} ``turbulent'' halos generated in merging clusters 
by turbulent re-acceleration of relativistic particles, and {\it (ii)} ``off-state'' halos generated by secondary electrons in more relaxed clusters. The maximum emitted frequency, $\nu_s,$\footnote{at larger frequencies the synchrotron spectrum of halos steepens. Following Cassano et al. (2010) we adopt the convention that RHs have spectral index $\alpha=1.9$ between $\nu_s/2.5$ and  $\nu_s$.} in turbulent halos depends ultimately on the turbulent energy budget that is available in the hosting cluster.
We follow a simplified approach based on two separate cluster radio-populations. Specifically, we assume that those clusters where turbulence is not sufficient to generate RHs emitting at the observing frequency, $\nu_o$, host ``off-state'' halos. 
We assume that presently observed giant RHs, those following the RH power -- cluster X-ray luminosity correlation (see Fig.\ref{Fig.Lx_P_sec}, left panel), are mainly driven by turbulent re-acceleration in merging clusters. Whereas the radio power of clusters hosting ``off-state'' halos is constrained by using limits derived for ``radio quiet'' galaxy clusters (see upper limits in Fig.\ref{Fig.Lx_P_sec}; Brunetti et al. 2007; Brown et al. 2011). Brown et al. (2011)  claimed the detection of  diffuse emission 
from ``off-state'' galaxy clusters by stacking SUMSS images of $\sim$ 100 clusters at a luminosity-level slightly below that constrained by the upper-limits in Fig.\ref{Fig.Lx_P_sec}. Optimistically in our modelling we shall assume that this is the level of the hadronically-induced "off state" halos.
 
We model the properties of the ``turbulent'' and ``off-state'' halos and their cosmic evolution by means of a Monte Carlo approach, which is based on the semi-analytic model of Lacey \& Cole (1993; \ie extended Press \& Schechter 1974) to describe the hierarchical process of formation of galaxy cluster dark matter halos. The merger history of a synthetic population of galaxy clusters is followed back in time and the generation of the turbulence in the ICM is estimated for each merger identified in the {\it merger trees}. It is assumed that turbulence is generated in the volume swept by the subcluster infalling into the main cluster and that a fixed fraction ($\sim 0.1-0.3$) of the PdV work done by this subcluster goes into MHD turbulence, which in turn becomes available for particle acceleration on Mpc-scale. We do not follow directly the process of magnetic field growth and amplification in the ICM, but this is anchored to the evolution of the host cluster, assuming a scaling between the mean rms magnetic field $\langle B \rangle$ and the virial cluster mass, $M_v$, $\langle B \rangle\propto M_v^b$, with $b\gtsim 0.5$ (\eg Dolag et al. 2002).

The most important expectation of turbulent re-acceleration scenarios is that the synchrotron spectrum of RHs (see Fig.\ref{Fig.Lx_P_sec}, right panel) should become gradually steeper above a frequency, $\nu_s$, that is determined by the competition between acceleration and energy losses and which is connected to the energetics of the merger events that generate the halos (\eg Fujita et al. 2003; Cassano \& Brunetti 2005). The frequency $\nu_s$ depends on the acceleration efficiency , $\chi$, and on $\langle B \rangle$,  as $\nu_s\propto \langle B \rangle\chi^2 /(\langle B \rangle^2+B_{cmb}^2)^2$ (\eg Cassano et al. 2006, 2010a)\footnote{$B_{cmb}=3.2 (1+z)^2 \mu$G is the equivalent magnetic field of the cosmic microwave background (CMB) radiation}. Monte Carlo simulations of cluster mergers that occur during the hierarchical process of cluster formation allow for evaluating $\chi$ from the estimated rate of turbulence-generation and the physical condition in the ICM, and consequently to explore the dependence of $\nu_s$ on cluster mass, redshift, and merger parameters in a statistical sample of synthetic clusters.
Consequently, in the adopted scenario the population of RHs is expected to be made of a complex mixture of sources with different spectra, with massive (and hot) clusters that have a tendency to generate halos with spectra, measured between two frequencies, that are flatter than those in less massive systems (Fig.\ref{Fig.Lx_P_sec}, right panel).

\noindent Contrary to turbulent halos, ``off-state'' halos are expected with power-law spectra with fairly similar slopes (Fig.\ref{Fig.Lx_P_sec}, right panel), independently of the cluster mass. Consequently, surveying the sky at different radio frequencies and with appropriate sensitivities allows to disentangle these two populations. 

In order to estimate the occurrence of RHs in surveys at different observing frequencies we assume that only those halos with $\nu_s \geq \nu_o$ can be observable, $\nu_o$ being the observing frequency. Energy arguments imply that 
giant RHs with $\nu_s \geq$ 1 GHz are generated in connection with the most energetic merger-events in the Universe. 
Only these mergers can generate enough turbulence on Mpc scales and potentially produce the acceleration rate that is necessary to maintain the 
relativistic electrons emitting at these frequencies (Cassano \& Brunetti 2005). In general, in this model the fraction of clusters with radio 
halos increases with the cluster mass, since more massive clusters are more turbulent (\eg Vazza et al. 2006; Hallman \& Jeltema 2011), and thus are more likely to host a turbulent RH. 
Present surveys carried out at $\nu_o \sim$ 1 GHz detect RHs only in the most massive and merging clusters (\eg Cassano et al. 2013). 

For similar energy arguments RH with lower values of $\nu_s$, or with ultra-steep radio spectra (USSRH\footnote{Operatively, in this paper we define USSRH as those halos with $\alpha>1.9$ between 250-600 MHz.}), must be more common, 
since they can be generated in connection with less energetic phenomena, \eg major mergers between less massive systems or minor mergers 
in massive systems, that are more common in the Universe.


\section{The luminosity function of radio halos}

\begin{figure}
\begin{center}
\includegraphics[width=0.8\textwidth]{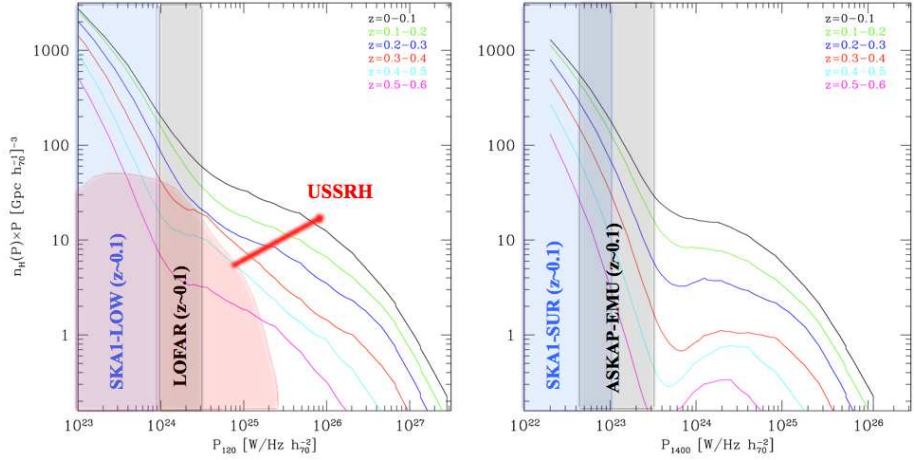}
\caption{Total RHLFs obtained by combining the contributions from ``turbulent'' and from ``off-state'' (hadronic) halos at 120 MHz (left panel) and 1400 MHz (right panel) are reported at different redshifts (see figure panels). The local RHLF of USSRH is also highlighted (shaded red region). The expected sensitivities to RHs of SKA1-LOW and SUR, LOFAR and EMU are shown at $z\sim0.1$ (shaded vertical regions).}
\label{Fig.RHLF}
\end{center}
\end{figure} 

The luminosity functions of turbulent RHs (RHLFs) with $\nu_s\geq \nu_0$ 
(\ie the expected number of halos per comoving volume and radio power ``observable'' at
frequency $\nu_0$) 
can be estimated by :

\begin{equation}
{dN_{H}(z)\over{dV\,dP(\nu_0)}}=
{dN_{H}(z)\over{dM\,dV}}\bigg/ {dP(\nu_0)\over dM}\,,
\label{RHLF}
\end{equation}

\noindent
where M is the virial cluster mass and $dN_{H}(z)/dM\,dV$ is the theoretical mass function of clusters hosting radio 
halos with $\nu_s \geq \nu_0$, that is obtained by combining Monte Carlo 
calculations of the fraction of clusters with RHs and 
the Press \& Schechter (PS) mass function of clusters (\eg Cassano et al. 2006).
Following Cassano et al. (2006) we estimate 
$dP(\nu_0)/dM$ from the correlation observed for giant RHs
between the 1.4 GHz radio power, $P(1.4)$, and the mass of the 
parent clusters (\eg Govoni et al. 2001; Cassano et al. 2006, Fig.\ref{Fig.Lx_P_sec}, left panel). 

\noindent The luminosity function of ``off-state'' halos is:

\begin{equation}
{{dN_H^{sec}(z,\nu_o)} \over {dV\,dP}}=
{{dN_H^{sec}(z,\nu_o)} \over {dV\,dM}}\times {{dM}\over{dP}}\,,
\label{RHLF_sec}
\end{equation}

\noindent 
where $dN_{H}(z)/dM\,dV$ is the massfunction of  clusters hosting ``off-state'' halos, which is given by:

\begin{equation}
{{dN_H^{sec}(z,\nu_o)} \over {dV\,dM}}=(1-f_{RH}(M,\nu_o)) 
\times {{dN_M^{cl}}\over {dV\,dM}}
\label{RHMF_sec}
\end{equation}

\noindent where $f_{RH}(M, \nu_o)$ is the fraction of clusters of mass $M$ with RHs 
due to turbulent re-acceleration (with $\nu_s \geq \nu_o$) and $dN_M^{cl}/dV\,dM$ is the cluster mass function. 
We derive $dM/dP$ from the expected relation between the radio luminosity of ``off-state'' halos 
and the mass (or $L_X$) of the host clusters (\eg Kushnir et al. 2009) that is slightly flatter than that of 
giant RHs.
In this Chapter, we adopt a simplified scaling in which
we consider a constant magnetic field $\langle B \rangle=3\mu$G and a normalization 
$P\approx 5\times 10^{23}$ Watt/Hz for $L_{X} = 10^{45}$ erg/sec 
(red line in Fig.\ref{Fig.Lx_P_sec}, left panel, see also Sect.3 in Cassano et al. 2012, for details on the adopted values for the
parameters). 

In Fig.~\ref{Fig.RHLF} we report the total RHLF, obtained by combining the contributions from ``turbulent'' RHs and from (purely hadronic) ``off-state'' halos and its redshift evolution (see figure panel). Under our assumptions, ``off-state'' halos dominate the RHLF at lower radio luminosities where the RHLF due to turbulent RHs flattens. The luminosity functions are derived for RHs with $\nu_s\ge 120$ MHz (left panel) and for halos with $\nu_s\ge 1000$ MHz (right panel). The contribution of USSRH to the local ($z\sim0.05$) RHLF at 120 MHz is also highlighted (shaded red region) showing that these systems contribute substantially to the RHLF at low radio powers, $P_{120}\sim 3-30 \times10^{23}$ W/Hz.


\section{Detecting giant radio halos in clusters}

To estimate the minimum flux of a RH (integrated over a scale of $\sim$ 1 Mpc) that can be detected in a survey we will consider two possible approaches:
{\it (i)} a brightness-based criterion and {\it (ii)} a flux-based criterion\footnote{Simulations of SKA1 observations of galaxy clusters are necessary to study the capability of the radio telescope to detect faint diffuse emission on different physical scales (see \eg Ferrari et al., this Volume).}.

\noindent 
The criterion based on a threshold in brightness guarantees that halos are detected in the images generated by the survey. We know that the typical brightness profiles of RH smoothly decrease with distance from the cluster center, so the outermost regions are very difficult to detect. Based on a sample of well-studied RHs, Brunetti et al. (2007) found that about half ($\sim 58\%$) of the total radio flux of a RH is contained in about half radius of the halo. What is important is thus the capability of the survey to detect at least the brightest central regions of the RH. The minimum flux, $f_{min}(z)$, of a 1 Mpc RH that can be detected in a survey can be obtained by requiring that the mean halo brightness within half halo radius is $\xi_1$ times ($\xi_1\approx2-3$) the noise level in the map:

\begin{equation}
f_{min}(z)\simeq 1.2 \times10^{-4} \xi_1\,
\Big(\frac{\mathrm{F_{rms}}}{10 \mu\mathrm{Jy}}\Big)
\Big(\frac{100\,\mathrm{arcsec}^2}{\theta_b^2}\Big)
\Big(\frac{\theta_{H}^2(z)}{\mathrm{arcsec}^2}\Big)
\,\, [\mathrm{mJy}]\,,
\label{fmin}
\end{equation}

\noindent
where $\theta_{H}(z)$ is the apparent angular radius of the halo at a given redshift in arcseconds, $\theta_b$ is the beam angular size in arcseconds and $F_{rms}$ is the rms noise per beam. 

\noindent 
A second possible approach to derive $f_{min}$ is to assume that the halo is detectable
when the integrated flux within $0.5\,\theta_H$ gives a signal to noise ratio $\xi_2$, \ie
$f_{min}(<0.5\,\theta_H)\simeq 0.58 f_{min}(<\theta_H) \simeq \xi_2\,\sqrt{N_b}\times F_{rms}$, where $N_{b}$ is the number of independent beam within $0.5\theta_H$, it follows: 

\begin{equation}
f_{min}(z)\simeq1.43\times10^{-3}\, \xi_2\,
\Big(\frac{\mathrm{F_{rms}}}{10 \mu\mathrm{Jy}}\Big)
\Big(\frac{10\,\mathrm{arcsec}}{\theta_b}\Big)
\Big(\frac{\theta_{H}(z)}{\mathrm{arcsec}}\Big)\, \, [\mathrm{mJy}]\,,
\label{fminhuub}
\end{equation}

\noindent Here we adopt $\xi_1=2$ and $\xi_2=7$ as reference values for calculations, these values are constrained through the detection of fake RHs in real {\it uv} radio data (NVSS, GMRT) and through the comparison between model expectations and observations (see Brunetti et al. 2007; Venturi et al. 2008; Cassano et al. 2012; for more details).
\begin{figure}
\begin{center}
\includegraphics[width=.4\textwidth]{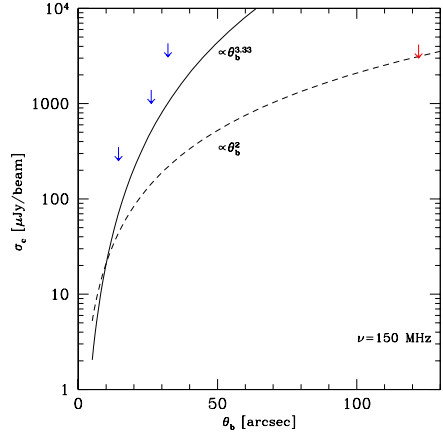}
\caption[]{Confusion noise, $\sigma_c$, as a function of $\theta_b$ as provided by Eq.~\ref{Eq.conf} (dashed line) and by Condon et al. (2012) using 3 GHz deep JVLA observations (solid line). The red arrow shows a measure of $\sigma_c$ at 150 MHz with WSRT (Bernardi et al. 2009), the blue arrows are measures of the rms noise at 150 MHz on deep GMRT fields (from left to right: Ishwara-Chandra et al. 2010; Intema et al. 2011; George \& Stevens 2008). Since these GMRT observations are not limited by confusion, the measured rms noise at different $\theta_b$ could be considered as upper limits on $\sigma_c$.}
\label{Fig.conf}
\end{center}
\end{figure}

In this Chapter we consider the radio surveys reported in Tab.\ref{Tab1}: the {\it LOFAR Tier 1} survey and SKA1-LOW survey at 120 MHz, and the EMU (ASKAP) and SKA1-SUR at 1400 MHz. 
\begin{table}
\begin{center}
\begin{tabular}{lcc}
\hline
\hline
configurations & rms & $\theta_b$ \\
			   & $\mu$Jy/beam & arcesc \\	
\hline
\hline
LOFAR (120 MHz) & 400 & 25 \\
SKA1-LOW (120 MHz) & 20 & 10 \\
\hline
EMU (1.4 GHz) & 13 & 15\\
SKA1-SUR (1.4 GHz) & 5 & 15\\
\hline
\hline	
\end{tabular}
\caption{Survey performance assumptions adopted in this Chapter.}
\label{Tab1}
\end{center}
\end{table}

It is important, especially for observations at low radio frequency, to evaluate the role of confusion. Indeed, as reported in Braun (2014), most of the SKA1-LOW continuum configurations would be limited by the confusion noise. The classical confusion noise due to faint radio sources have been estimated at GHz frequencies (Condon 1987):

\begin{equation}
\Big(\frac{\mathrm{\sigma_{c}}}{\mathrm{mJy\,beam^{-1}}}\Big)\approx 0.2 \Big(\frac{\mathrm{\nu}}{\mathrm{GHz}}\Big)^{-0.7}\Big(\frac{\mathrm{\theta_b}}{\mathrm{arcmin}}\Big)^2
\label{Eq.conf}
\end{equation}

As illustrated in Fig.\ref{Fig.conf}, this is roughly consistent with recent results by Condon et al. (2012; obtained with the JVLA at ~3 GHz and with 8 arcsec resolution, continuum line) for resolutions $\sim5-15$ arcsec, while for larger beams Eq.~\ref{Eq.conf} provides a more reliable estimate of $\sigma_c$ in comparison with some observed values (arrows). For the {\it LOFAR Tier 1} survey (\eg R\"ottgering 2010) we will assume $F_{rms}$=0.4 mJy/beam and $\theta_b\sim25$ arcsec \footnote{this is a conservative assumption, since pointed LOFAR observations obtained using configurations similar to the  {\it Tier 1} survey reached rms values $\sim 0.1$  mJy/beam at $\sim 5$ arcsec resolution and $\sim0.2-0.3$ mJy/beam at $\sim 20-30$ arcsec (van Weeren et al. in prep.)}; according to Eq.\ref{Eq.conf} this is still far from being confusion-limited. On the other hand, considering Eq.\ref{Eq.conf} a SKA1-LOW continuum survey with $F_{rms}=20 \mu$Jy/beam, $\theta_b=10$ arcesc is already confusion limited. 

At higher frequency, corresponding to the SKA1-SUR band, ASKAP can perform cluster science improving the sensitivity and resolution with respect to current radio surveys at $\sim 1$ GHz (such as the NVSS).
 One important key project of ASKAP will be EMU, the ``Evolutionary Map of the Universe'' (Norris et al. 2011), an all-sky continuum survey with a sensitivity of 10  $\mu$Jy/beam 
 and an angular resolution of 10 arcsec. For the EMU survey we consider a configuration with $\theta_b=15$ arcsec and $rms=13 \mu$Jy/beam. 
On the other hand SKA1-SUR could be able to perform all-sky surveys in 1-2 GHz band with $rms=2 \mu$Jy/beam and $\theta_b=2$ arcsec. Such a survey would be far from confusion and can be used to make source-subtracted images, that can be tapered up to a resolution of 15 arcsec to increase the sensitivity to the extended emission. In this case we can consider a survey at 1.4 GHz with $rms=5 \mu$Jy/beam and $\theta_b=15$ arcsec.\footnote{Our estimates of telescope sensitivities to diffuse emission are probably conservative, because it might be possible to mitigate confusion by subtracting compact sources (see \eg Vernstrom et al 2014).} 
The equivalent of EMU in the northern sky will be the WODAN (Westerbork Observations of the Deep APERTIF Northern-Sky) survey (R\"ottgering et al. 2011); the expectations that we will derive in next Sections for EMU can be used also for WODAN.

\begin{figure}
\begin{center}
\includegraphics[width=6cm,height=6cm]{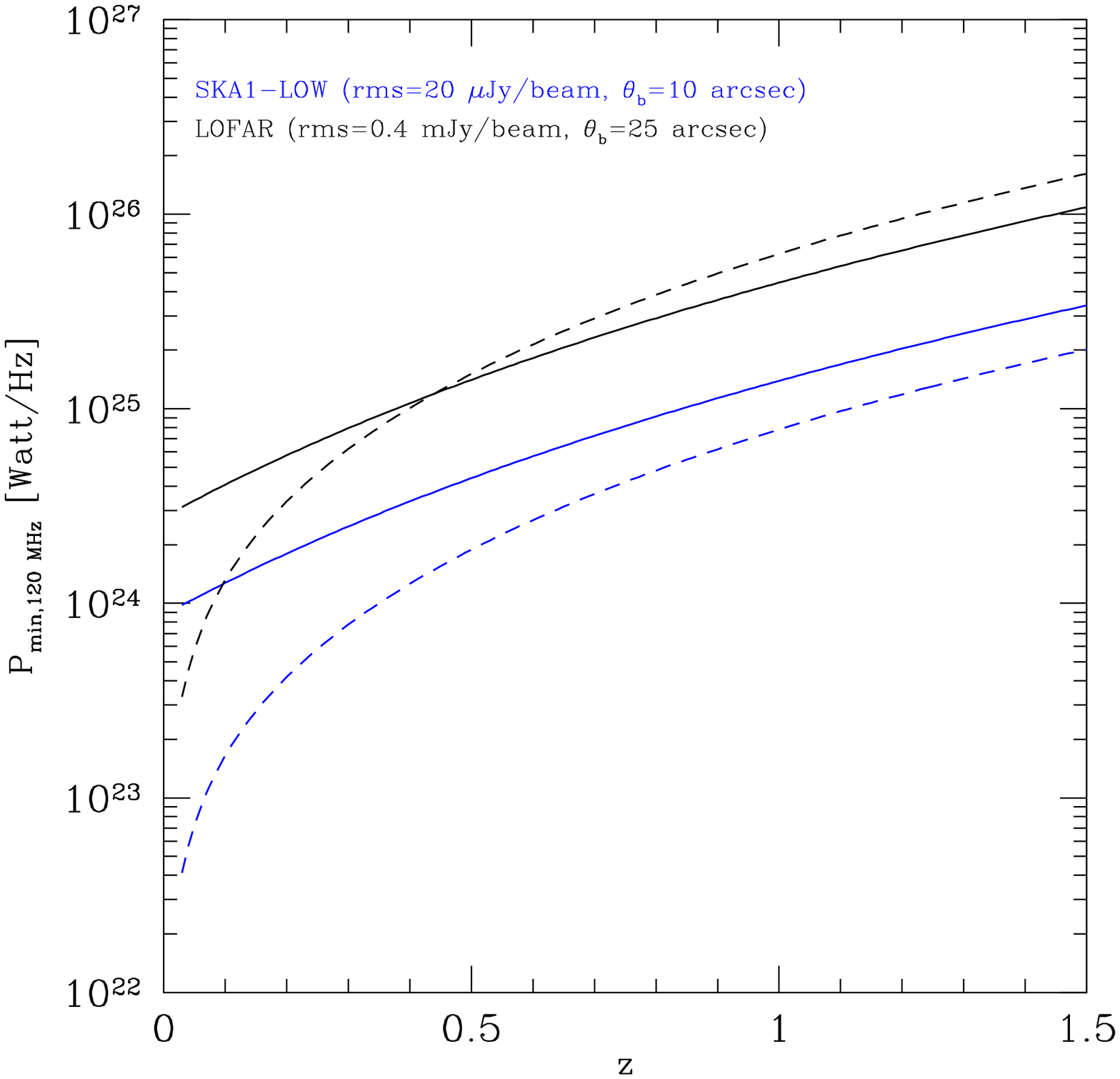}
\includegraphics[width=6cm,height=6cm]{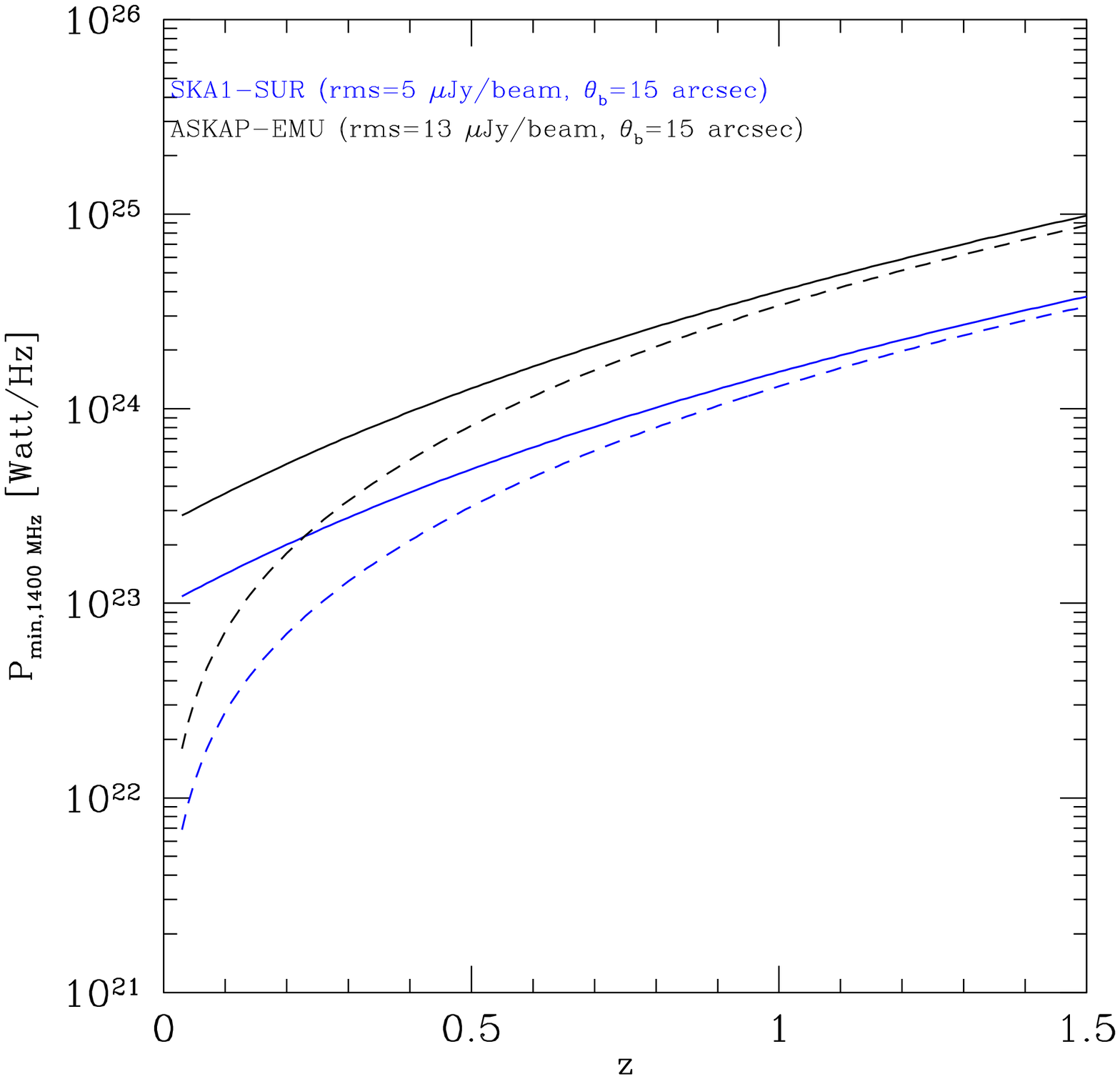}
\caption[]{Minimum power of RHs detectable at 120 MHz (left panel) and at 1400 MHz (right panel) in different radio surveys (see figure panels) as a function of redshift. The minimum radio power has been computed according to Eqs.\ref{fmin} (dashed lines) and \ref{fminhuub} (solid lines).}
\label{Lrmin_z}
\end{center}
\end{figure}

Fig.~\ref{Lrmin_z} shows the resulting minimum power of RHs detectable in the considered surveys.

\section{Number of Radio Halos in surveys with SKA1-LOW and SUR}

 \begin{figure}
\begin{center}
\includegraphics[width=0.7\textwidth]{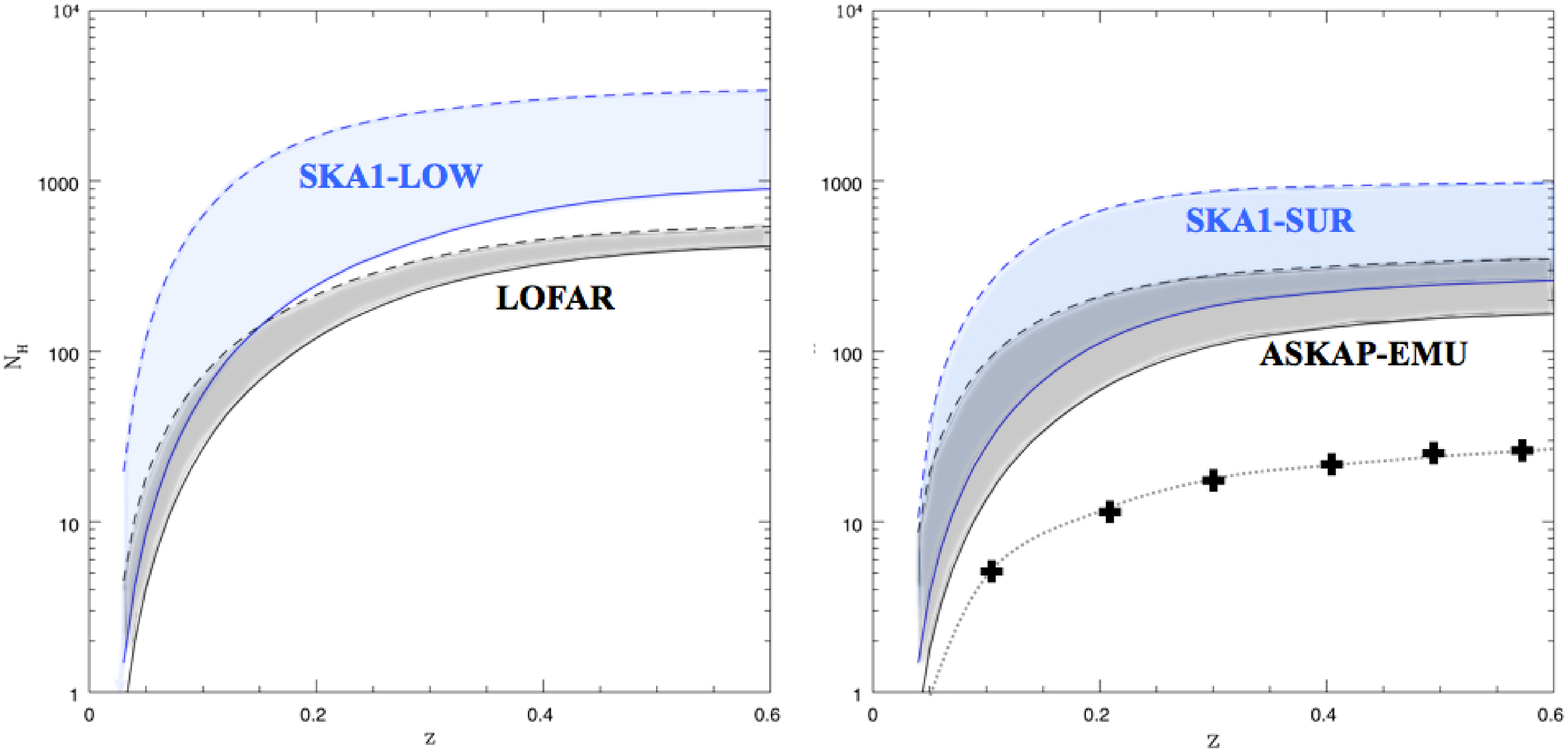}
\includegraphics[width=0.7\textwidth]{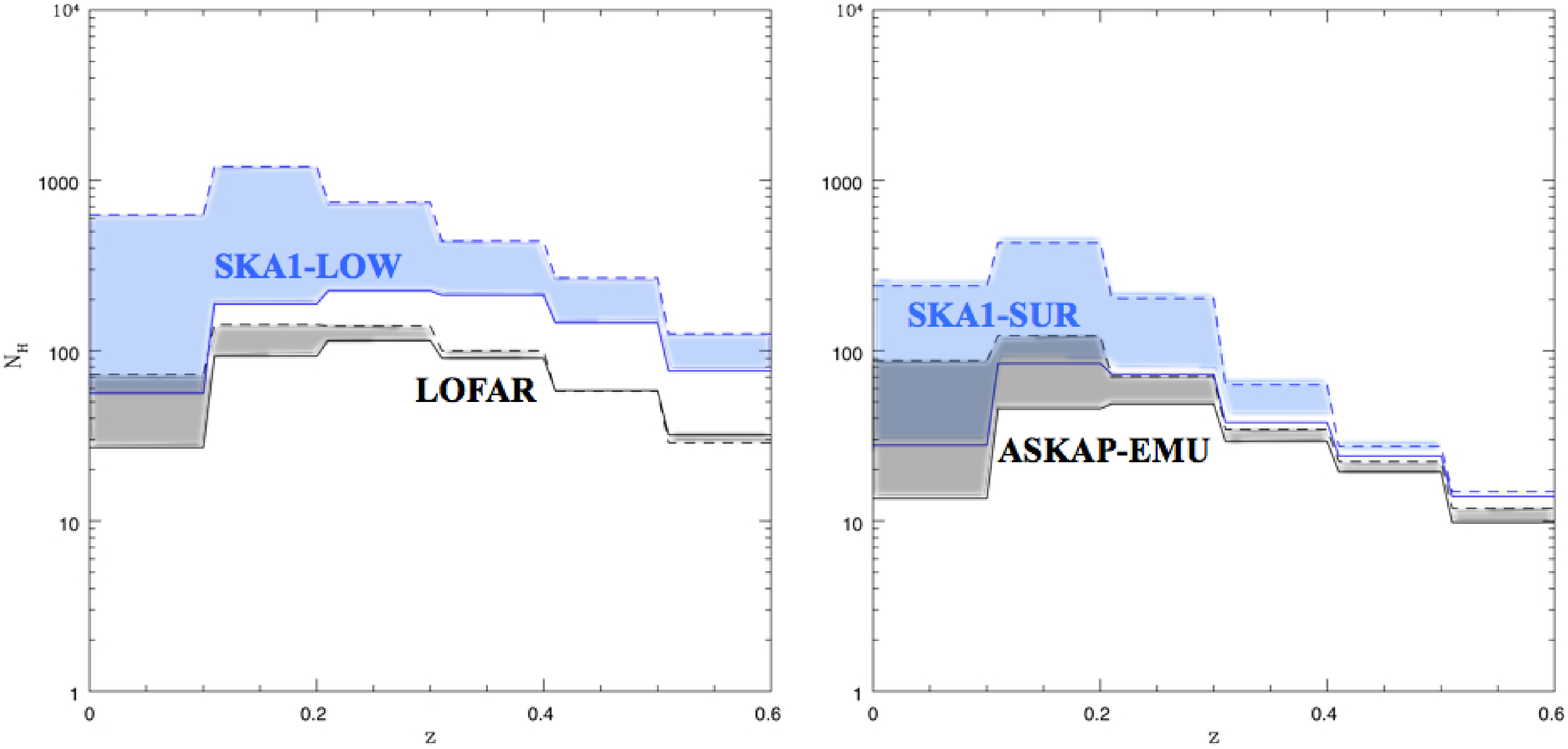}
\caption[]{{\it Upper Panels}: expected integral number (all-sky) of RHs as a function of redshift at 120 MHz (left panel) and 1400 MHz (right panel). Black crosses show the integral number of RHs observed so far at $\sim 1$ GHz. {\it Lower Panels}: expected number of RH (all-sky) in redshift intervals at 120 MHz (left panel) and 1400 MHz (right panel). In all plots, the shaded regions show the ranges obtained considering the two methods (Eq.\ref{fmin} and Eq.\ref{fminhuub}) to derive the RH detection limit in the considered survey (see figure panels).}
\label{Fig.NH}
\end{center}
\end{figure} 
        
The number of RHs with $flux\geq f_{min}(z)$ in the redshift interval, $\Delta z=z_2-z_1$, can be obtained by combining the RHLF ($dN_H(z)/dP(\nu_o)dV$) and $f_{min}(z)$:

\begin{equation}
N_{H}^{\Delta_z}=\int_{z=z_1}^{z=z_2}dz' ({{dV}\over{dz'}})
\int_{P_{min}(f_{min}^{*},z')}{{dN_H(P(\nu_o),z')}\over{dP(\nu_o)\,dV}}\,
dP(\nu_o)
\label{Eq.RHNC}
\end{equation}

In Fig.\ref{Fig.NH} we show the all-sky number of RHs expected in the LOFAR (black) and SKA1-LOW (blue) surveys (left panels) and in the EMU  (black) and SKA1-SUR  (blue) surveys (right panel). 
We consider both giant RHs that originate from turbulent re-acceleration in merging clusters and ``off-state'' halos. We consider the flux limit derived according to Eq.~\ref{fmin} with $\xi_1=2$ (solid lines) and that obtained by Eq.~\ref{fminhuub} with $\xi_2=7$ (dashed lines).

\noindent
As a general consideration, we note that with the current design SKA1-LOW is much more efficient in the detection of RHs than SKA1-SUR. 
Given the typical RHLF (Fig.\ref{Fig.RHLF}) ``off-state'' halos are expected to contribute significantly to the total number of RHs in both low frequency and mid frequency surveys at lower redshift ($z<0.3$). LOFAR would mainly detect RH generated in turbulent merging clusters, while SKA1-LOW and SKA1-SUR could start to test {\it for the very first time} the presence of hadronic halos in relaxed clusters. 
In both cases, LOFAR and SKA1-LOW would be able to detect USSRH, with the number of these sources being larger in SKA1-LOW surveys thanks to the better sensitivity. The detection of a number of these USSRH is a powerful test for models (Brunetti et al. 2008).

SKA1-LOW and SKA1-SUR could be able to detect up to $\sim2600$ and $\sim750$ halos, respectively, on $3\pi$ sr and out to $z\sim0.6$. We note that this difference is due both to the better sensitivity of SKA1-LOW to the detection of diffuse steep-spectrum cluster scale emission and to the presence of USSRH, which would be detectable preferentially at lower radio frequency. For comparison, the maximum number of halos detectable by LOFAR and EMU would be $\sim 400$ and $\sim 260$, respectively. At both low and mid frequency SKA1 promises a tangible gain in the RH detection with respect to precursors and pathfinders. A major step can be already obtained with the early phase of SKA1-LOW (50\% of SKA1-LOW sensitivity) that could be able to detect more than 1000 RH. The SKA2 at low frequency would be sensitive to RHs that are two times less powerful than those visible by SKA1-LOW, with the potential to increase by an order of magnitude the number of RH detected by SKA1, because the number counts of ``off-state'' halos are expected to be very steep.

\section{Radio Halos to detect galaxy clusters in radio surveys}

\begin{figure}
\begin{center}
\includegraphics[width=0.8\textwidth]{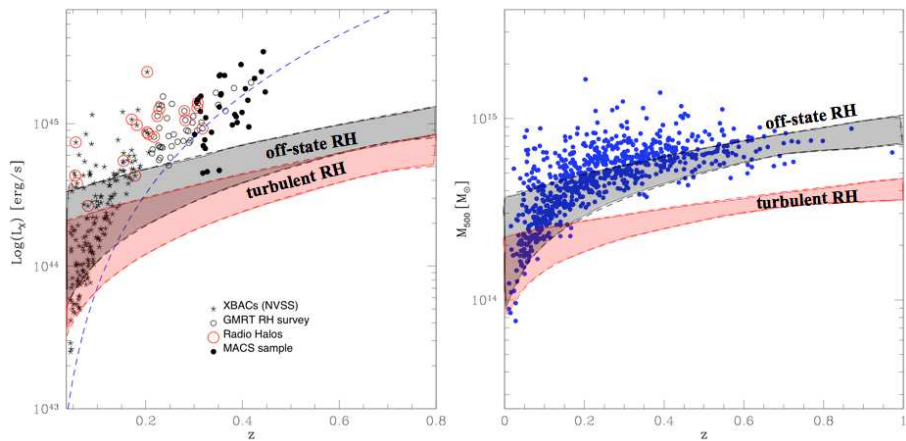}
\caption[]{{\it Left Panel}: distribution of clusters in the $L_X-z$ plane (see legend in the figure panel). The sensitivity limit of the Reflex sample (blue dashed line) is reported. {\it Right Panel}: distribution of clusters detected by {\it Planck} in the $M_{500}-z$ plane. In both panels, the shaded regions show the minimum X-ray luminosity and mass of clusters with ``turbulent'' halos (red regions) and with ``off-state'' halos (black regions) detectable in the SKA1-LOW survey.}
\label{Fig.RH_clusters}
\end{center}
\end{figure} 

In this Section we compare the potential of the SKA1-LOW survey to discover galaxy clusters using RHs with that of other surveys in the X-ray and SZ.
By using Eq.~\ref{fmin} and Eq.~\ref{fminhuub} we derive the minimum power of giant RH detectable in the SKA1-LOW survey as a function of redshift. Then, using the scaling relations $P_{1.4}-L_{X}$ (\eg Cassano et al. 2006) and $P_{1.4}-M_{500}$ (\eg Basu 2012; Cassano et al. 2013) for ``turbulent'' RH and ``off-state'' halos separately, we derive the minimum X-ray luminosity and mass of clusters that can be detected in SKA1-LOW as a function of z. Fig.\ref{Fig.RH_clusters}, left panel shows that
SKA1-LOW would be more powerful in the detection of galaxy clusters with respect to present X-ray surveys, especially at $z\ge0.2$. In Fig.\ref{Fig.RH_clusters}, right panel, we show the distribution of {\it Planck} clusters in the $M_{500}-z$ plane. Because of the very steep relation between the RH power and the cluster mass ($P_{1.4}\propto M_{500}^{3.7}$, Cassano et al. 2013), the cluster selection function provided by SKA1-LOW in the $M_{500}-z$ plane is relatively flat, becoming competitive with the Planck SZ cluster survey in the detection of high-z clusters. A preliminary analysis suggest that SKA1-LOW offers the unique opportunity to detect high-z clusters through their diffuse radio emission in the form of both ``turbulent''  and ``off-state'' halos. Clearly at these redshifts the possibility to separate the truly diffuse emission from the contribution of discrete sources becomes challenging. Although simulations of SKA1 observations are needed (see Ferrari et al., this Volume), we note that the $10''$ resolution of SKA1-LOW correspond to a physical scale of $\sim 80$ kpc at $z\sim1$, suggesting that a first order estimate of the contribution of the discrete sources will be possible.

\section{Radio halos at high-z: caution}

\noindent In the previous Section we showed that SKA1-LOW is potentially powerful in the detection of high-z clusters through the detection of their RH emission. On the other hand, only a few massive high-z ($z>0.5-0.6$) clusters are know to host giant RH (\eg Bonafede et al. 2009, 2012; van Weeren et al. 2009, 2014; Lindner et al. 2014). This number 
starts to increase only recently thanks to the detection of high-z massive clusters via SZ-based cluster surveys (with SPT, Planck, etc.). 

Theoretically, the generation of high-z RH is challenging due to the increase with z of the Inverse Compton losses of the radio emitting electrons ($dE/dt \propto(1+z)^4$). Present models, based on semi-analytical calculations of cluster formation and turbulence generation, are only able to make trustworthy expectations for moderate-z ($z<0.5-0.6$) clusters. This is because the PS-based Monte Carlo method starts to underproduce the number and merging rate of high-z clusters. This means that in the adopted formalism the formation of massive clusters is delayed (massive clusters start to form at lower z) with respect to what is observed in cosmological simulations or predicted by more refined semi-analytic models (\eg Giocoli 2012, 2013).
This implies that the expectations presented for RH at $z\sim 0.5-0.6$ (Fig.\ref{Fig.NH}) should be taken as lower limits.
More refined semi-analytical models to describe the formation and evolution of galaxy clusters (\eg Giocoli 2012, 2013) combined with our formalism for non-thermal components will allow to obtain more reliable expectations at higher redshift (Cassano et al in prep).


\end{document}